\documentclass[pre,showpacs,floatfix,preprint,superscriptaddress,byrevtex,aps]{revtex4-1}
\usepackage[english]{babel}
\usepackage[utf8x]{inputenc}
\usepackage{amsmath,amssymb,amstext}
\usepackage{mathrsfs}
\usepackage{graphicx}
\usepackage{subfigure}
\usepackage{color}

\newcommand{\ket}[1]{\left\vert #1\right\rangle}
\newcommand{\bra}[1]{\left\langle#1\right\vert}
\newcommand{\braket}[2]{{\left\langle \vphantom{#1 #2} #1 \,\right|
   \left.\hspace{-0.15em} \vphantom{#1 #2} #2\right\rangle}}
\newcommand{\matel}[3]{{\left\langle \vphantom{#1 #2 #3} #1 \,\right\vert \left.
  \hspace{-0.15em} \vphantom{#1 #2 #3} #2 \,\right\vert \left.
  \hspace{-0.15em} \vphantom{#1 #2 #3} #3\right\rangle}}
\newcommand{\eps}{\varepsilon}

\newcommand{\be}{\begin{equation}}
\newcommand{\ee}{\end{equation}}
\newcommand{\RR}{\mathbb{R}}
\newcommand{\NN}{\mathbb{N}}

\newcommand{\ZZ}{\mathbb{Z}}

\newcommand{\bea}{\begin{eqnarray}}
\newcommand{\eea}{\end{eqnarray}}
\newcommand{\sign}{\mathrm{sign}}

\newcommand{\cD}{\mathcal{D}}

\begin{document}

\title{Some properties of the one-dimensional L\'{e}vy crystal}
\author{B. A. Stickler}
\email{benjamin.stickler@uni-graz.at}
\affiliation{Institute of Physics,
 Karl-Franzens Universit\"{a}t Graz, A-8010 Graz, Austria}
 
\begin{abstract}
We introduce and discuss the one-dimensional L\'{e}vy crystal as a probable candidate for an experimentally accessible realization of space fractional quantum mechanics (SFQM) in a condensed matter environment. The discretization of the space fractional Schr\"{o}dinger equation with the help of shifted Gr\"{u}nwald-Letnikov derivatives delivers a straight-forward route to define the L\'{e}vy crystal of order $\alpha \in (1,2]$. As key ingredients for its experimental identification we study the dispersion relation as well as the density of states for arbitrary $\alpha \in (1,2]$. It is demonstrated that in the limit of small wavenumbers all interesting properties of continuous space SFQM are recovered, while for $\alpha \to 2$ the well-established nearest neighbor one-dimensional tight binding chain arises.
\end{abstract}

\maketitle
 
\section{Introduction}

Space fractional quantum mechanics (SFQM), as introduced by N. Laskin \cite{laskin2000a,laskin2000b,laskin2002}, is a natural generalization of standard quantum mechanics which arises when the Brownian trajectories in Feynman path integrals are replaced by L\'{e}vy flights. The classical L\'{e}vy flight is a stochastic processes which, in one dimension, is described by a jump length probability density function (pdf) of the form \cite{metzler2000,dubkov2008}
\be \label{eq:levy}
p_\alpha(x) \propto \frac{1}{\vert x \vert^{\alpha + 1}}, \qquad \text{for } \vert x \vert \to \infty.
\ee
where $\alpha \in (0,2]$ is referred to as the L\'{e}vy index. Although there are numerous applications of classical L\'{e}vy flights \cite{metzler2000} such as, for instance, the description of particle trajectories in a rotating flow \cite{solomon93} or the traveling behavior of humans \cite{brockmann06}, to my knowledge, to this day no experimental realization or observation of SFQM has been reported. While the current literature seems to concentrate on more mathematical aspects of the theory \cite{jeng10,jianping13}, it is still an open question for which systems manifestations of SFQM are to be expected. It is the aim of this paper to address this shortcoming and to point the route towards an experimental realization of SFQM.

In order to position the current work in an appropriate context, let us briefly survey the basic notions of SFQM. The one-dimensional (1D) space fractional Schr\"{o}dinger equation reads \cite{laskin2000a,laskin2000b}
\be \label{eq:fracschr}
D_\alpha \hat P^\alpha \ket{\psi} + \hat V \ket{\psi} = E \ket{\psi},
\ee
where $D_\alpha \in \RR$ [$D_2 = 1 / (2m)$] is a constant, $\hat P^\alpha$ is the $\alpha$-th power of the momentum operator, $\hat V$ is the potential operator, and $\ket{\psi}$ is the eigenstate pertaining to eigenenergy $E$. The requirement of the first moments of the L\'{e}vy process with pdf \eqref{eq:levy} to exist, constrains $\alpha$ to $\alpha \in (1,2]$ \cite{laskin2002}. The position space representation of the $\alpha$-th power of the momentum operator is given by \cite{laskin2002}
\be \label{eq:momop}
\matel{x}{\hat P^\alpha}{\psi} = - \hbar^\alpha \cD_{\vert x \vert}^\alpha \psi(x),
\ee
where $\cD_{\vert x \vert}^\alpha$ is the Riesz fractional derivative operator of order $\alpha$ \cite{kilbas2006} and $\hbar$ is the reduced Planck constant. For $\alpha = 2$ the Riesz fractional derivative is equivalent to the standard second order derivative \cite{kilbas2006}. The free particle solution of the space fractional Schr\"{o}dinger equation \eqref{eq:fracschr} is easily determined to be of the form \cite{laskin2000a,laskin2000b}
\be \label{eq:freep}
\psi(x) = \exp( i k x) \qquad \text{with} \qquad E = D_\alpha \hbar^\alpha \vert k \vert^\alpha,
\ee
i.e. the dispersion is proportional to $\vert k \vert^\alpha$. Hence, in the limit $\alpha \to 2$ we obtain a parabolic dispersion as, for instance, for conduction electrons near the band minimum \cite{ibach} and for $\alpha \to 1$, the dispersion shows the behavior $\vert k \vert$ as, for instance, a 1D acoustic phonon band for $k \to 0$ \cite{madelung}. The solution of the fractional Schr\"{o}dinger equation in other, more complex situations appears to be rather difficult due to the nonlocality of Eq. \eqref{eq:fracschr} \cite{jeng10}. It is an important property of the space fractional Schr\"{o}dinger equation that it is not possible to solve the equation locally and to obtain a global solution from matching conditions as it is usual in standard quantum mechanics. Numeric solutions for the 1D particle in a box have, for instance, have been presented in \cite{herrmann13}.

These open questions make it, apparently, rather difficult to search for possible realizations of SFQM. For instance, Lenzi {\it et al.} \cite{lenzi2008} determined the specific heat of non-crystalline solids for low temperatures with the help of a fractional generalization of the thermodynamic Green's function, however, their fits to experimental data resulted in values $\alpha > 2$. On the other hand, in Ref. \cite{stickler11c} the semiclassical evolution equation resulting from Eq. \eqref{eq:fracschr} has been investigated. It was left as an open problem which particular form of the Wigner transform has to be employed and, therefore, which semiclassical behavior of the free flight term is to be expected.

Within this paper we suggest a new route toward an experimental identification of SFQM. In particular, we discuss a probable candidate for the realization of SFQM in a condensed matter environment by introducing a 1D infinite range tight binding chain which we shall refer to as the 1D L\'{e}vy crystal of order $\alpha \in (1,2]$ for reasons of convenience. This model emerges from the mapping of the Hamiltonian \eqref{eq:fracschr} onto a 1D lattice and, thus, represents a natural generalization of the well established 1D nearest neighbor tight binding chain ($\alpha = 2$). It has to be emphasized that such a model can only be regarded as an idealization and that, from this point of view, SFQM has to be understood as an effective theory which may be employed under certain conditions. Furthermore, the investigation of the L\'{e}vy crystal allows to resolve the dilemma of ambiguous Wigner transforms raised in \cite{stickler11c}, at least within this context.

This paper is structured as follows: in Sec. \ref{sec:crys} we define the L\'{e}vy crystal, in Sec. \ref{sec:disp} we determine its dispersion relation as well as its density of states (DOS). Moreover, we discuss an interesting interpretation of this model and briefly investigate the dilemma raised in \cite{stickler11c}. Finally, in Sec. \ref{sec:conc} the work is summarized.

\section{The 1D L\'{e}vy Crystal} \label{sec:crys}

\begin{figure}
 \centering
 \includegraphics[width = 120mm]{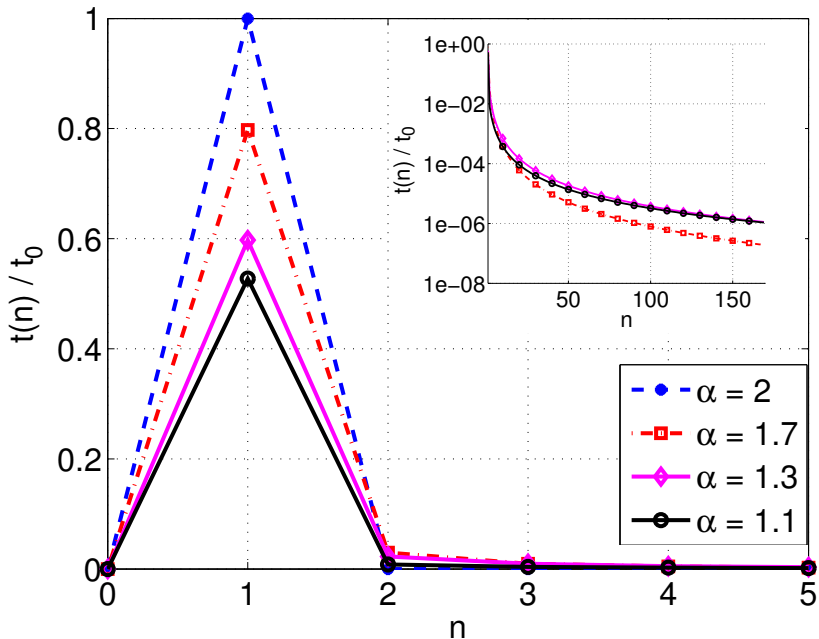}
 \caption{(color online)The normalized hopping parameters $t(n) / t_0$ as a function of the distance $n$ vs $n$ for different values of $\alpha \in (1,2]$. In the inset we display the same curves on a logarithmic scale.} \label{fig:hoppings}
\end{figure}

We start with a closer inspection of the space-fractional Schr\"{o}dinger equation \eqref{eq:fracschr} by expressing the Riesz fractional derivative operator $\cD_{\vert x \vert}^\alpha$ as \cite{gorenflo98}
\be \label{eq:24}
\cD^\alpha_{\vert x \vert} = - \frac{1}{2 \cos \left (\frac{ \alpha \pi}{2} \right )} \left ( I^{-\alpha}_+ + I^{-\alpha}_- \right ),
\ee
where $I_\pm^{-\alpha}$ can be written with the help of shifted Gr\"{u}nwald-Letnikov derivatives \cite{gorenflo98} as
\be \label{eq:25}
I^{-\alpha}_\pm \psi(x) = \lim_{a \to 0} \frac{1}{a^\alpha} \sum_{n = 0}^\infty (-1)^n \binom{\alpha}{n} \psi [ x \mp (n-1)a ].
\ee
for sufficiently well behaved functions $\psi(x)$. Here, $\binom{\alpha}{n}$ is the generalized binomial coefficient defined by
\be \label{eq:3}
\binom{\alpha}{n} = \frac{\Gamma(\alpha+1)}{\Gamma(n+1) \Gamma(\alpha-n+1)},
\ee
with $\Gamma(\cdot )$ the $\Gamma$ function. In order to arrive at a tight binding model we follow the procedure outlined by S. Datta \cite{datta} and replace the exact momentum operator \eqref{eq:24} by its discretized version. We regard equally spaced grid-points $x_\ell = \ell a$, where $a >0$ is the lattice constant and $\ell \in\ZZ$. The position space element for a particular gridpoint $x_\ell$ of the kinetic term of Eq. \eqref{eq:fracschr} acting on $\ket{\psi}$ is, therefore, replaced by
\begin{widetext}
\bea \label{eq:kinterm}
\matel{x_\ell}{- D_\alpha \hat P^\alpha}{\psi} \to \matel{x_\ell}{- D_\alpha \hat P^\alpha_a}{\psi} := \frac{t_0}{2} \sum_{ n = 0}^\infty  (-1)^n \binom{\alpha}{n} \left \{ \psi[x_\ell + ( n - 1)a ] + \psi [ x_\ell - (n - 1) a ] \right \},
\eea
\end{widetext}
where we defined the hopping amplitude $t_0$:
\be
t_0 = \frac{D_\alpha \hbar^\alpha}{a^\alpha \cos \left ( \frac{\alpha \pi}{2} \right ) }.
\ee
In a first step we rewrite Eq. \eqref{eq:kinterm}:
\begin{widetext}
\bea
\matel{x_\ell}{- D_\alpha \hat P^\alpha_a}{\psi} & = & \frac{t_0}{2} \sum_{n \neq 0} (-1)^{n+1}  \binom{\alpha}{n+1} \psi(x_\ell + n a) + \frac{t_0}{2} \left [ \psi(x_\ell - a) + \psi(x_\ell + a) \right ] - \alpha t_0 \psi(x_\ell).
\eea
\end{widetext}
Then we assume that the potential operator $\hat V$ is diagonal in position space and periodic with periodicity $a$, i.e. $V(x_\ell) \equiv U$ for all $\ell \in \ZZ$. This suggests the replacement
\bea \label{eq:repl}
\matel{x_\ell}{-D_\alpha \hat P^\alpha + \hat V}{\psi} & \to & \eps \psi(x_\ell) + \sum_{n \neq 0} t(n) \psi(x_\ell + na),
\eea
where we defined the onsite energy
\be
\eps = U - \alpha t_0,
\ee
together with hopping parameters
\be \label{eq:2}
t(n) = \frac{t_0}{2} \left [ (-1)^{\vert n \vert +1} \binom{\alpha}{\vert n \vert +1} + \delta_{\vert n \vert,1} \right ],
\ee 
for $\alpha \in (1,2]$, $n \neq 0$ and $\delta_{nm}$ is the Kronecker $\delta$. The hopping parameters \eqref{eq:2} are illustrated in Fig. \ref{fig:hoppings} for different values of $\alpha \in (1,2]$. Let us briefly discuss this particular form of the hopping parameters $t(n)$. First of all, we note that the constraint $\alpha \in (1,2]$ ensures that $t(n) / t_0 \geq 0$ for all $n \in \NN$. Furthermore, for $\alpha = 2$, the hopping parameters \eqref{eq:2} reduce to
\be
t(n) = t_0 \delta_{\vert n \vert,1},
\ee
i.e. they account only for nearest neighbor interaction and, therefore, give rise to a 1D nearest neighbor tight binding chain. This is entirely consistent with the above discretization since the replacement \eqref{eq:kinterm} is equivalent to the finite difference approximation of the second order derivative for $\alpha = 2$ \cite{datta}. Moreover, we note that the hopping parameters obey
\bea \label{eq:4}
\frac{1}{t_0} \sum_{n = -\infty}^\infty t(n) & = & \alpha,
\eea
where we employed that \cite{gorenflo98}
\be \label{eq:5}
\sum_{n = 0}^\infty (-1)^n \binom{\alpha}{n}z^n = (1 - z)^\alpha,
\ee
for $\vert z \vert \leq 1$ and we defined $t(0) \equiv 0$.

In a final step we identify the replacement \eqref{eq:repl} with the definition of a tight binding Hamiltonian
\be \label{eq:1}
H = \eps \sum_n \ket{n} \bra{n} + \sum_{nm} t( m - n ) \ket{n} \bra{m},
\ee
where $\ket{n}$ are the basis-kets which are centered at position $x_n$ and form an orthonormal basis, i.e. $\braket{n}{m} = \delta_{nm}$. The tight binding Hamiltonian \eqref{eq:1} defines the L\'{e}vy crystal for $\alpha \in (1,2]$ with onsite energy $\eps$ and hopping amplitude $t_0$ if $t(n)$ is given by Eq. \eqref{eq:2}. We are now in the position to study some of the properties of this system. It is important to remark that an alternative discretization of the integrals \eqref{eq:25} would have resulted in an alternative tight binding Hamiltonian \eqref{eq:1}. However, the main aspects of the following discussion remain unchanged if an alternative discretization would have been chosen. Moreover, we chose the particular version \eqref{eq:25} because it is a straight forward generalization of the 1D monoatomic nearest neighbor tight binding model and, thus, the electronic structure will have only one maximum/minimum in the first Brioullin zone.

\section{Dispersion Relation and Density of States} \label{sec:disp}

The electronic structure as well as the DOS of the L\'{e}vy crystal might be of central significance for an experimental identification of SFQM. In a first step, we solve the stationary Schr\"{o}dinger equation for the Hamiltonian \eqref{eq:1} following the standard procedure \cite{economou}. We express the wavefunction $\ket{\psi}$ as a linear combination of the localized orbitals $\ket{\ell}$, i.e.
\be \label{eq:6}
\ket{\psi} = \sum_\ell c_\ell(k) \ket{\ell},
\ee
then, we employ Bloch's theorem
\be \label{eq:7}
c_{\ell}(k) = \exp \left [ i k a (\ell - \ell') \right ] c_{\ell'}(k),
\ee
where we introduce the wavenumber $k \in \left [-\frac{\pi}{a}, \frac{\pi}{a} \right]$ and insert Eq. \eqref{eq:6} into Eq. \eqref{eq:1}.  Together with Eq. \eqref{eq:7} we obtain
\bea \label{eq:8}
H \ket{\psi} & = & \eps \ket{\psi} + \sum_{n \ell} t( \ell - n) c_\ell \ket{n} \notag \\
  & = & \left \{ \eps + 2 \mathrm{Re} \left [ \sum_{\ell \geq 1} t(\ell) \exp \left ( i k a \ell \right ) \right ] \right \} \ket{\psi},
\eea
where $\mathrm{Re}(\cdot)$ denotes the real part. It is important to note that Bloch's theorem, Eq. \eqref{eq:7}, is valid because the L\'{e}vy crystal is infinitely extended. It is, therefore, comparable to the plane wave solution of the one-dimensional space fractional Schr\"{o}dinger equation \eqref{eq:fracschr} \cite{laskin2000a,laskin2000b}.

\begin{figure}
 \centering
 \includegraphics[width = 80mm]{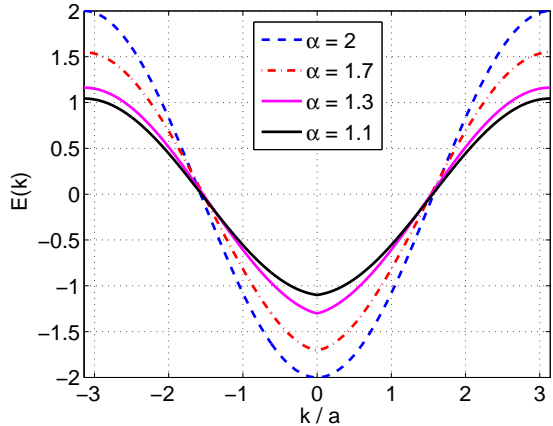}
 \caption{(color online)Dispersion relation $E_\alpha(k)$ vs $k$ for different values of $\alpha \in (1,2]$ and the particular choice $\eps = 0$ and $t_0 = -1$.} \label{fig:dispersion}
\end{figure}

With the help of Eq. \eqref{eq:5} we can express the sum for $\alpha \in (1,2]$ as
\begin{widetext}
\be \label{eq:10}
\sum_{\ell \geq 1} t(\ell) \exp ( i k a \ell) = \frac{t_0}{2} \exp( - i k a) \left [ 1 - \exp ( i k a) \right ]^\alpha + i t_0 \sin ( ka ) + \frac{\alpha t_0}{2}.
\ee
Hence, the dispersion relation of the 1D L\'{e}vy crystal is described by
\be \label{eq:11}
E_\alpha(k) = \eps + \alpha t_0 + t_0 \mathrm{Re} \left \{ \exp(- i k a) \left [ 1 - \exp(ika) \right]^\alpha \right \} \quad \text{ for } \alpha \in (1,2].
\ee
This illustrated in Fig. \ref{fig:dispersion}. In the particular case that $\alpha = 2$ the dispersion reduces to
\be \label{eq:12}
E_2(k ) = \eps + 2 t_0 \cos(ka),
\ee
which is well known from the 1D nearest neighbor tight binding chain \cite{datta,economou}. Next, we transform the dispersion relation \eqref{eq:11} into a more convenient form. We note that
\bea \label{eq:len}
\vert \exp ( - i ka ) \left [ 1 - \exp (i k a ) \right ]^\alpha \vert & = & 2^{\frac{\alpha}{2}} \left [ 1 - \cos(ka) \right ]^{\frac{\alpha}{2}} = 2^{\alpha} \left \vert \sin \left ( \frac{ka}{2} \right ) \right \vert^{\alpha}.
\eea
Moreover, we rewrite
\bea
\arg \left [ 1 - \exp ( i k a) \right ] & = & - \arctan \left ( \frac{\sin ( k a )}{1 - \cos ( k a ) } \right )= - \arctan \left [ \cot \left ( \frac{ka}{2} \right ) \right ] = \frac{ka}{2} - \frac{\pi}{2} \sign (k),
\eea
and, hence,
\bea \label{eq:arg}
\arg \left \{ \exp ( - i k a ) \left [ 1 - \exp ( i k a) \right ]^\alpha \right \} = - ka + \frac{\alpha}{2} \left [ ka - \pi \sign (k) \right ].
\eea
Combining Eqs. \eqref{eq:len} and \eqref{eq:arg} finally allows us to rewrite the dispersion \eqref{eq:11} as
\bea \label{eq:dispersion2}
E_\alpha(k) & = & \eps + \alpha t_0 + 2^{\alpha} t_0  \left \vert \sin \left ( \frac{ka}{2} \right ) \right \vert^{\alpha} \cos \left [ ka \left (1 - \frac{\alpha}{2} \right )  + \frac{\alpha \pi}{2} \sign (k) \right ] \notag \\
\eea
\end{widetext}

In the limit $a \to 0$ or $k \to 0$ we expect, due to our definition of the L\'{e}vy crystal, to obtain a dispersion which resembles the free particle space fractional Schr\"{o}dinger equation \eqref{eq:freep}. Indeed, we obtain from Eq. \eqref{eq:dispersion2}
\be \label{eq:18}
E_\alpha(k) \approx \eps + \alpha t_0 + t_0 (a \vert k \vert )^\alpha \cos \left ( \frac{\alpha \pi}{2} \right ),
\ee
for small $\vert k \vert$. Hence, as expected, the dispersion \eqref{eq:11} of the L\'{e}vy crystal is non-analytic in $k$ at $k = 0$, indicating an infinite band velocity. The question arises, whether such a model can be of any interest for real-world applications. Obviously, the particular form of Eq. \eqref{eq:18} is the result of the infinite sum \eqref{eq:10} evaluated with the help of Eq. \eqref{eq:5}. More precisely, it is a result of the asymptotic behavior of the hopping parameters \eqref{eq:2}. Let us briefly investigate how the dispersion \eqref{eq:10} is changed if the hoppings are truncated to finite range interactions, i.e. if we set $t(n) = 0$ for all $n > M$, where $M \in \NN $ is left arbitrary. Then, the dispersion \eqref{eq:11} takes on the form
\bea \label{eq:dispersion3}
E_\alpha(k) & = & \eps + 2 \mathrm{Re} \left [ \sum_{\ell = 1}^M t(\ell) \exp ( i k a \ell) \right  ]. \notag \\
 & = & \eps + 2 \sum_{\ell = 1}^M t( \ell) \cos ( k a \ell) \notag \\
 & \approx & \eps + 2 \sum_{\ell = 1}^M t(\ell) \left [ 1 - \frac{(k a \ell )^2}{2} \right ],
\eea
i.e. the dispersion is analytic and parabolic for $k \to 0$ for all $M < \infty$. Hence, the L\'{e}vy crystal can be regarded as an effective model for a crystal which possesses a dispersion proportional to $\vert k \vert^\alpha$ in the {\em vicinity} of $k = 0$.

Let us turn our attention to some further properties of the dispersion relation of the L\'{e}vy crystal. We restrict our discussion to the case $t_0 < 0$ for reasons of simplicity. From the dispersion relation \eqref{eq:11} we observe that the band minimum is located at $k = 0$ and takes on the value
\be \label{eq:13}
E_\alpha(k = 0) = \eps + \alpha t_0,
\ee
while the band maximum at $k = \pm \frac{\pi}{a}$ is given by
\be \label{eq:14}
E_\alpha \left ( k = \pm \frac{\pi}{a} \right ) = \eps - \left ( 2^\alpha - \alpha \right ) t_0.
\ee
Please note that $2^\alpha - \alpha > 0$ for all $\alpha \in (1,2]$. Hence, the total bandwidth $\delta_\alpha$ is given by
\bea \label{eq:delta}
\delta_\alpha : & = & \left \vert E_\alpha( k = 0) -  E_\alpha \left ( k = \pm \frac{\pi}{a} \right ) \right \vert \notag \\
 & = & 2^\alpha \vert t_0 \vert,
\eea
i.e. it increases with increasing $\alpha$. Interestingly, we have
\bea \label{eq:middle}
E_\alpha \left ( k = \pm \frac{\pi}{2 a} \right ) & = & \eps + \alpha t_0  + t_0 \mathrm{Re} \left [ i (1 + i )^\alpha \right ] \notag \\
 &= & \eps - t_0 \left [ 2^{\frac{\alpha}{2}} \sin \left ( \frac{\alpha \pi}{4} \right ) - \alpha \right ],
\eea
and we note that
\be \label{eq:17}
\left [ 2^{\frac{\alpha}{2}} \sin \left ( \frac{\alpha \pi}{4} \right ) - \alpha \right ] \geq 0,
\ee
where the equal sign applies to $\alpha = 2$. Hence, the point $k = \pm \frac{\pi}{2 a}$ does not coincide with the inflection point of the dispersion for $\alpha < 2$. This already indicates that the DOS cannot be symmetric in $E$.

\begin{figure*}
 \centering
 \subfigure[$~\alpha = 2$]{
\includegraphics[width = 79mm]{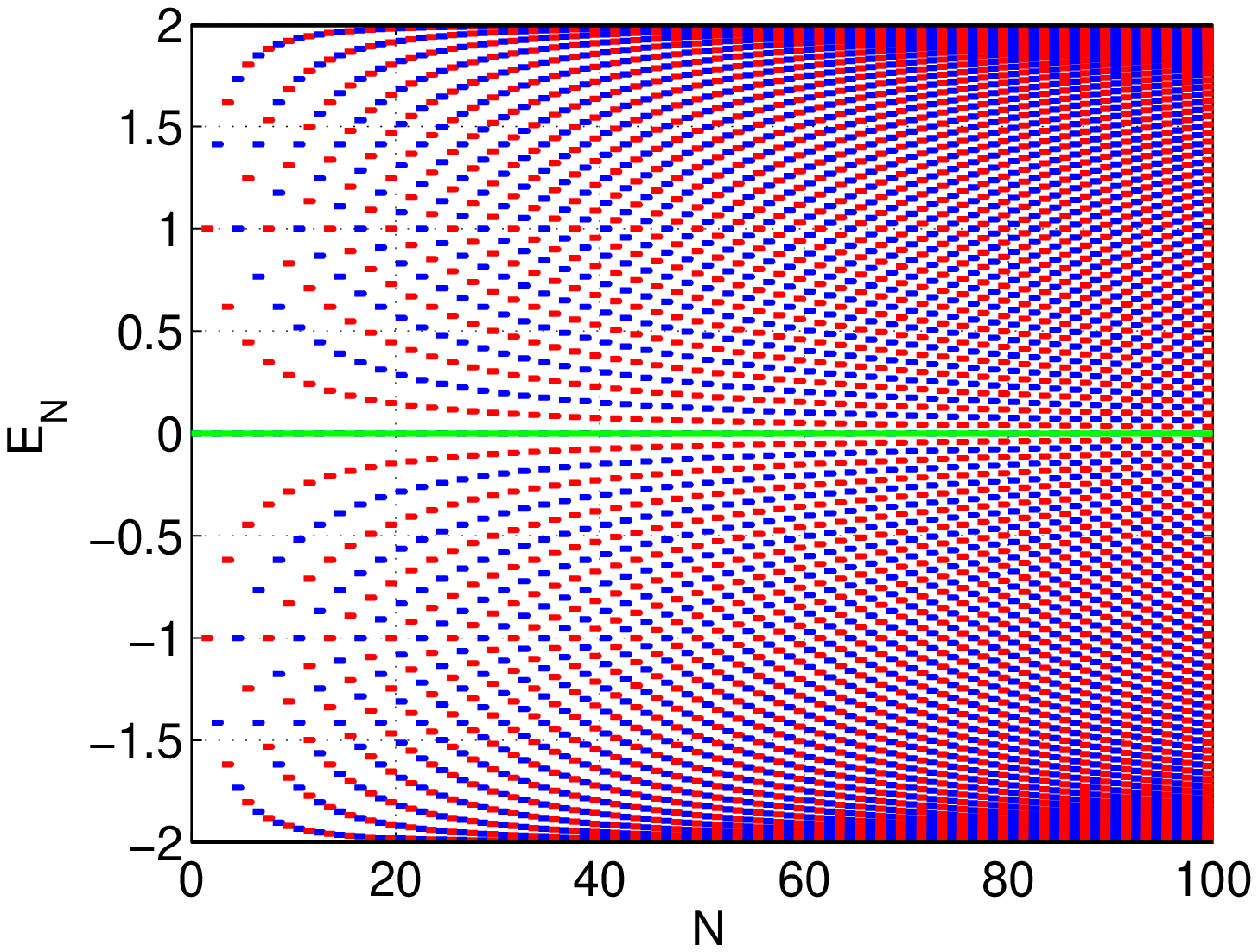}
}
\subfigure[$~\alpha = 1.7$]{
\includegraphics[width = 79mm]{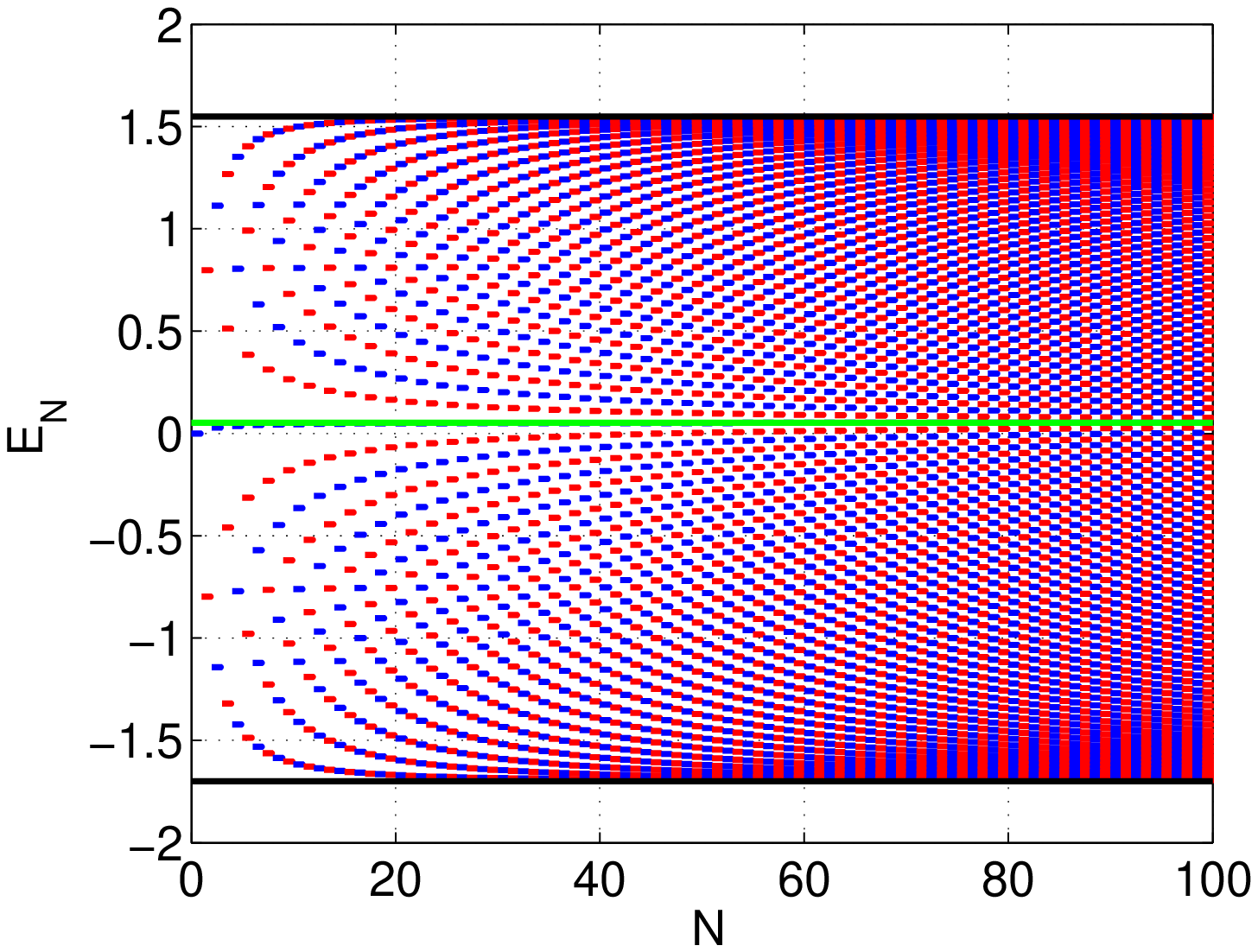}
}
\subfigure[$~\alpha = 1.3$]{
\includegraphics[width = 79mm]{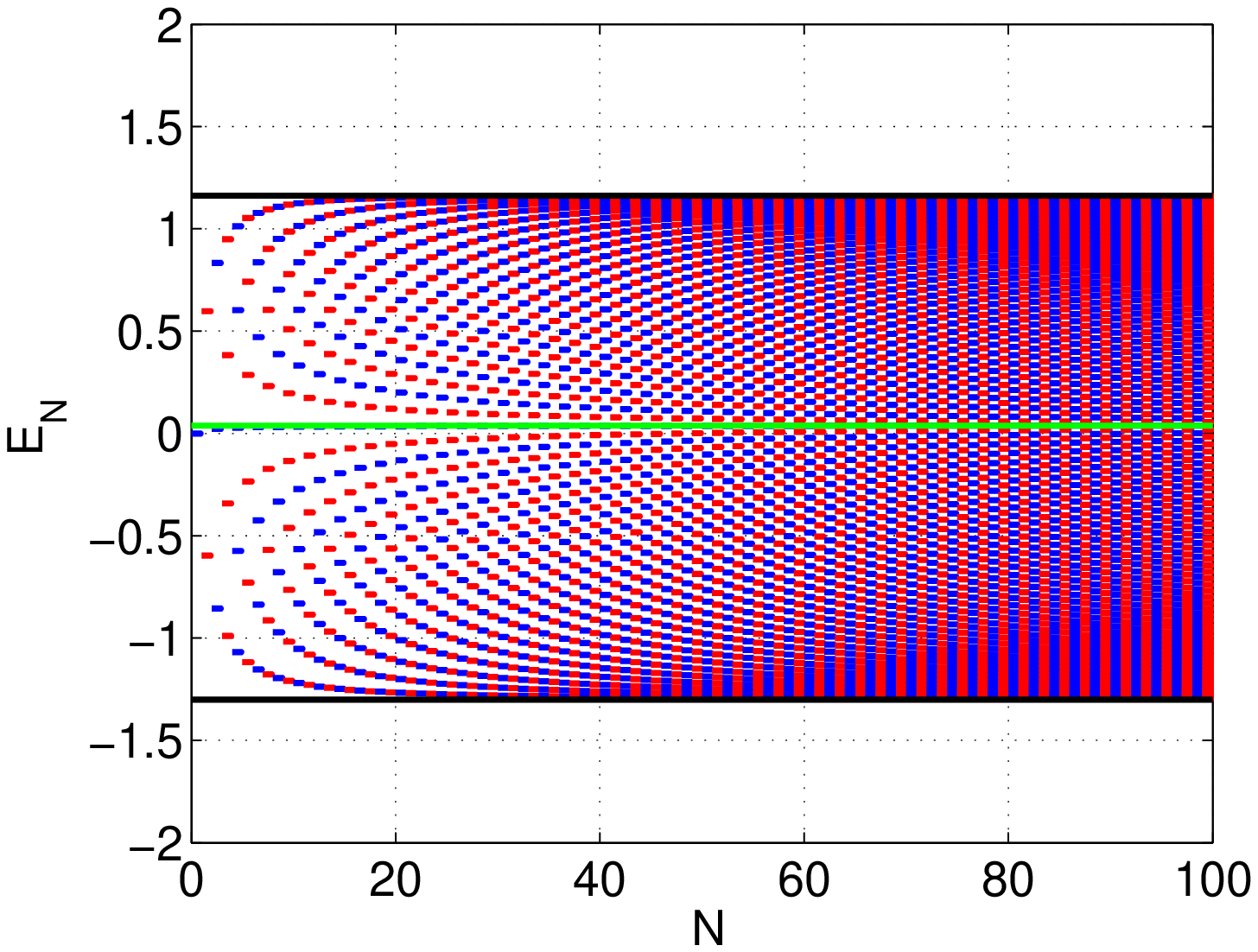}
}
\subfigure[$~\alpha = 1.1$]{
\includegraphics[width = 79mm]{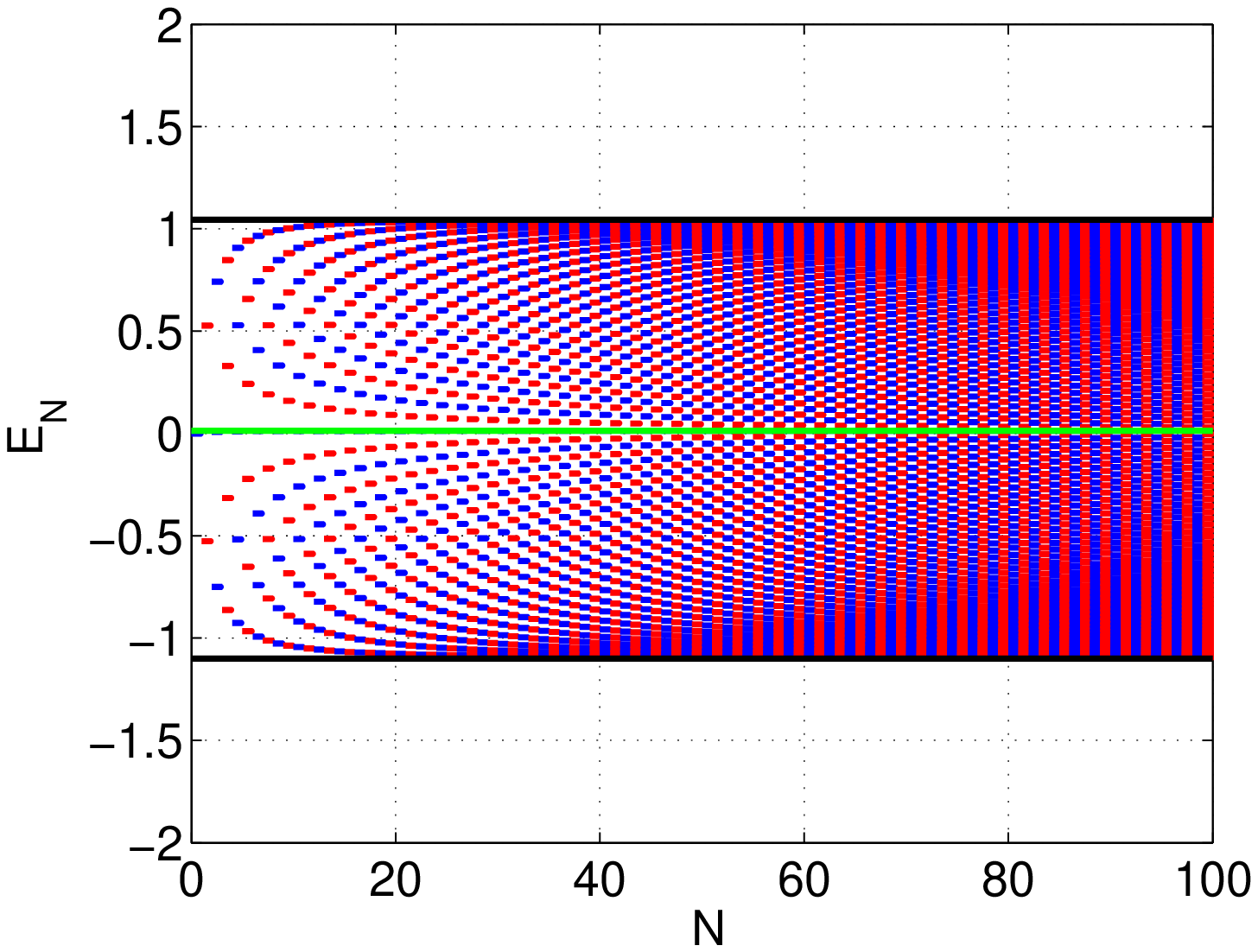}
}
\caption{(color online)Energy levels $E_N$ for different $\alpha \in (1,2]$ vs $N$ together with the characteristic points for $N \to \infty$ at $k = 0, \pm \frac{\pi}{2 a}, \pm \frac{\pi}{a}$. Moreover, we chose $\eps = 0$ and $t_0 = -1$.} \label{fig:enlevels}
\end{figure*}

Of course, we may also regard a L\'{e}vy crystal of finite length, i.e. in the Hamiltonian \eqref{eq:1} $n$ is restricted to $n = -N, \ldots, N$ with $N \in \NN$. In this case, Bloch's theorem \eqref{eq:7} cannot be applied and an analytic solution of the problem is not straight forward. In fact, this situation is comparable to the solution of the space fractional Schr\"{o}dinger equation \eqref{eq:fracschr} on a finite domain, as in the case of the one-dimensional particle in a box \cite{jeng10,herrmann13}. In Figs. \ref{fig:enlevels}(a) to \ref{fig:enlevels}(d) we illustrate the numerically obtained energy levels of the L\'{e}vy crystal of finite length for $\alpha = 2$, $\alpha = 1.7$, $\alpha = 1.3$ and $\alpha = 1.1$, respectively, and for different values of $N$ together with the characteristic points of the dispersion, Eqs. \eqref{eq:13}, \eqref{eq:14} and \eqref{eq:middle}. From this graphs we observe that the DOS increases with decreasing $\alpha$ due to the reduced bandwidth. 
Moreover, as stated above, the DOS is larger for $E>E \left ( \frac{\pi}{2 a} \right )$ than for $E < E \left ( \frac{\pi}{2 a} \right )$ since $E\left ( \frac{\pi}{2 a} \right )$ is slightly shifted to positive energies according to Eq. \eqref{eq:middle} and there has to be the same number of states above this energy and below.

\begin{figure}
 \centering
 \includegraphics[width = 80mm]{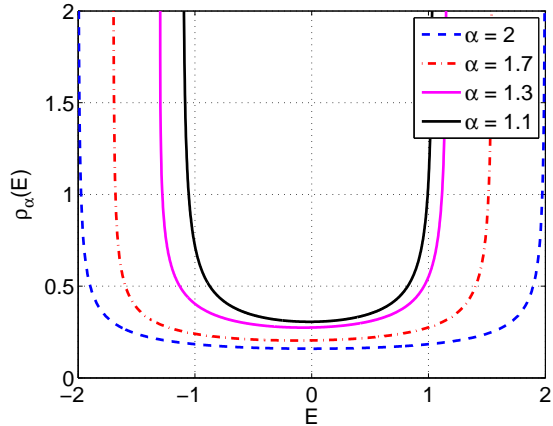}
 \caption{(color online)Density of states $\rho_\alpha(E)$ of the L\'{e}vy crystal for different $\alpha \in (1,2]$ vs $E$. Again, we chose $\eps = 0$ together with $t_0 = -1$.} \label{fig:dos}
\end{figure}

Let us derive an analytic expression for the DOS of the 1D L\'{e}vy crystal. The DOS $\rho_\alpha(E)$ of the 1D dispersion $E_\alpha(k)$ reads \cite{ibach}
\be
\rho_\alpha(E) = \left. \frac{1}{\pi} \frac{1}{ \vert \partial_k E_\alpha(k) \vert} \right \vert_{k = k(E)},
\ee
where $k(E)$ is the inverse dispersion and $\partial_k$ denotes the partial derivative with respect to $k$. With the help of Eq. \eqref{eq:dispersion2} we obtain for $k \neq 0$
\begin{widetext}
\bea
\left \vert \partial_k E_\alpha (k) \right \vert & = & \left \vert \alpha \frac{a 2^\alpha}{2}  t_0  \sign(k) \left \vert \sin \left ( \frac{ka}{2} \right ) \right \vert^{\alpha - 1} \cos \left ( \frac{ka}{2} \right ) \cos \left [ ka \left ( 1 - \frac{\alpha}{2} \right ) + \frac{\alpha \pi}{2} \sign(k) \right ] \right. \notag \\
 &  & \left. - t_0 2^\alpha a \left ( 1 - \frac{\alpha}{2} \right ) \left \vert \sin \left ( \frac{ka}{2} \right ) \right \vert^\alpha \sin \left [ ka \left ( 1 - \frac{\alpha}{2} \right ) + \frac{\alpha \pi}{2} \sign(k) \right ] \right \vert \notag \\
 & = & a 2^ \alpha \left \vert t_0 \sin \left ( \frac{ka}{2} \right ) \right \vert^{\alpha - 1} \left \vert \frac{\alpha}{2} \cos \left ( \frac{ka}{2} \right ) \cos \left [ ka \left ( 1 - \frac{\alpha}{2} \right ) + \frac{\alpha \pi}{2} \sign(k) \right ] \right. \notag \\
 &  & \left. - \left ( 1 - \frac{\alpha}{2} \right ) \sin \left ( \frac{ka}{2} \right ) \sin \left [ ka \left ( 1 - \frac{\alpha}{2} \right ) + \frac{\alpha \pi}{2} \sign(k) \right ] \right \vert.
\eea
\end{widetext}
For $k = 0$, the first derivative of the dispersion, which is essentially the band velocity, diverges. Finally, we have to compute the inverse of the electronic structure \eqref{eq:dispersion2}. First of all, we note the time reversal invariance of the dispersion $E_\alpha(k) = E_\alpha(-k)$ and, for $t_0 < 0$, that $\eps - \alpha \vert t_0 \vert \leq E_\alpha(k) \leq \eps + (2^\alpha - \alpha ) \vert t_0 \vert$. We rewrite Eq. \eqref{eq:dispersion2} as
\be \label{eq:omega}
\omega := \frac{E - \eps + \alpha \vert t_0 \vert}{ 2^\alpha \vert t_0 \vert} = - \left \vert \sin \left ( \frac{\phi}{2} \right ) \right \vert^\alpha \cos \left [ \phi \left ( 1  - \frac{\alpha}{2} \right ) + \frac{\alpha \pi}{2} \right ],
\ee
where $0 \leq \omega \leq 1$ and $\phi = ka$. We restrict the above equation to $\phi \in [0, \pi ]$ due to time reversal invariance of the band structure. Eq. \eqref{eq:omega} is solved numerically for different $\omega \in [0,1]$ in order to obtain $k(E)$. In Fig. \ref{fig:dos} we present the DOS for $\alpha = 2$, $\alpha = 1.7$, $\alpha = 1.3$ and $\alpha = 1.1$, respectively. Again, we observe that the DOS at $E = 0$ increases with decreasing $\alpha$. This results from the reduced band width, see Eq. \eqref{eq:delta}, Fig. \ref{fig:dispersion} and Fig. \ref{fig:enlevels}. Moreover, the DOS is not symmetric with respect to $E = 0$ for $\alpha \neq 2$. This phenomenon is due to the shifted center of energy Eq. \eqref{eq:middle} and can already be observed in Figs. \ref{fig:enlevels}.

The two main characteristica, namely that the dispersion is proportional to $\vert k \vert^\alpha$ and the asymmetry of the DOS might be of great interest to identify the L\'{e}vy crystal in a condensed matter environment.

Before concluding let us briefly focus onto another interesting point: The formulation of the space fractional Schr\"{o}dinger equation \eqref{eq:fracschr} as the low wavenumber limit of a TB Hamiltonian allows for an interesting interpretation. In the general condensed matter case, the Hamiltonian is of the form $\hat H = \hat T + \sum_n \hat V_n$ where $\hat T$ denotes the standard kinetic energy operator and $\hat V_n$ is the potential operator pertaining to unit cell $n$. If we employ the ansatz $\ket{\psi} = \sum_n c_n \ket{n}$ and calculate the matrix elements $\matel{\ell}{\hat H}{\Psi}$ we have to cope with terms of the form $\matel{\ell}{\hat V_m}{n}$. At this point it is common to employ two different simplifications \cite{slater54}: (a) the three-center approximation states that matrix elements $\matel{\ell}{\hat V_m}{n}$ are only non-zero if at least two of the three indices $\ell, m,n$ coincide and, (b), the $M$-th nearest neighbor approximation states that $\matel{\ell}{\hat V_n}{n}$ 
is only non-zero if $\vert n - \ell \vert \leq M$. This offers a very interesting interpretation of the L\'{e}vy crystal Hamiltonian \eqref{eq:1}: Again, we employ the three center approximation and identify $\eps = \matel{n}{\hat T}{n} + \matel{n}{\hat V_n}{n}$ and $\matel{m}{\hat V_m}{n} = t(n - m)$. If $\matel{m}{\hat V_m}{n} = t(m - n)$ follows the particular behavior \eqref{eq:2}, the space fractional Schr\"{o}dinger equation might be an appropriate model for small values of $k$. Moreover, we deduce that the Wigner transform II discussed in \cite{stickler11c} might be the adequate one for this model. This follows from the fact that the kinetic term of the semiclassical evolution equation for a solid state system is of the form $\nabla_k E(k) \cdot \nabla_x w(x,k,t)$, with $E(k)$ the dispersion and $w(x,k,t)$ the semiclassical distribution function \cite{jauhobook}. Inserting for $E(k)$ the dispersion of the L\'{e}vy crystal $E_\alpha(k)$ gives for small $k$ the equation resulting from Wigner transform 
II in Ref. \cite{stickler11c}.

\section{Conclusion} \label{sec:conc}

We defined and investigated some properties of the 1D L\'{e}vy crystal. Its definition was particularly motivated by the quest for a possible realization of SFQM in a solid state environment. Of course, the L\'{e}vy crystal has to be regarded as an idealized model comparable to the 1D monoatomic nearest neighbor tight binding model, which follows in the limit $\alpha = 2$. Let us briefly review the main steps of our discussion.

The definition of the L\'{e}vy crystal is based on the discretization of the space fractional Schr\"{o}dinger equation with L\'{e}vy index $\alpha$ with the help of shifted Gr\"{u}nwald-Letnikov derivatives \cite{gorenflo98} on an equally spaced grid. In analogy to the introduction of the 1D nearest neighbor tight binding chain \cite{datta}, the grid-points are interpreted as lattice points, which finally defines the L\'{e}vy crystal of order $\alpha$. Hence, in this picture the L\'{e}vy crystal may be regarded as an effective model for a 1D crystal if the overlap integrals $\matel{m}{\hat V_m}{n}$ show the characteristic distance dependence of $t(n - m)$ given by Eq. \eqref{eq:2}. The asymptotic behavior of the hopping parameters gives rise to a dispersion relation $E_\alpha(k) \propto \vert k \vert^\alpha$ in the limits $k \to 0$ or $a \to 0$. Furthermore, for $\alpha \to 2$ the well known nearest neighbor tight binding model arises \cite{economou}. These considerations allow to identify the Wigner 
transform II of Ref. \cite{stickler11c} as the correct one for this particular case. 
The DOS $\rho_\alpha(k)$ of the L\'{e}vy crystal for $\alpha\neq 2$ is no longer symmetric in $E$. Interestingly, we obtain a higher density of states for $E > E_\alpha \left ( \frac{\pi}{2a} \right )$ since the central state located at $E_\alpha \left ( k = \pm \frac{\pi}{2 a} \right )$ is shifted to higher energies, see Fig. \ref{fig:enlevels}. Moreover, for decreasing $\alpha$ the bandwidth decreases, which forces the overall DOS to increase.

\acknowledgments{I thank E. Schachinger for his interest in this work and for carefully reading this manuscript.}

\bibliographystyle{unsrt}

\end{document}